\documentstyle[eqsecnum,aps,epsfig]{revtex}
\newcommand{\be}{\begin{equation}}
\newcommand{\ee}{\end{equation}}
\newcommand{\bea}{\begin{eqnarray}}
\newcommand{\eea}{\end{eqnarray}}

\newcommand{\lton}{\mathrel{\lower.9ex
                  \hbox{$\stackrel{\displaystyle <}{\sim}$}}}

\begin{document}
\title{How the HBT-Puzzle at RHIC might dissipate}
\author{Adrian Dumitru}
\address{Physics Department, Brookhaven National Laboratory, P.O.\ Box 5000, 
Upton, NY 11973, USA}
\date{\today}
\maketitle   
\begin{abstract}
I compute the first-order corrections to two-particle Bose-Einstein
correlations
due to deviations from equilibrium. Based on that result I
argue that for nearly perfect fluids, the HBT radii ``freeze out'' much
later than the single-inclusive distribution.
Therefore, to prevent a big longitudinal homogeneity length,
the QGP presumably should hadronize not into a nearly perfect fluid of
hadrons but rather into a very dissipative one. This could be achieved
by a hadronization phase transition away from equilibrium.\\[1cm]
\end{abstract}


Pion interferometry has become a powerful tool for studying the size and
duration of particle production in high-energy collisions, from $e^+e^-$ over
$pp$ or $p\bar{p}$ to heavy ions like $Au+Au$ or
$Pb+Pb$~\cite{HBT,BertschPratt}.
For the case of nuclear collisions, the interest mainly focuses on the
possible transient formation of a deconfined state of matter.
This could affect the size of the
region from where the pions are emitted as well as
the time for particle production. In particular, it was hoped that a 
strong first-order QCD phase transition would lead to long lifetimes of
the particle source~\cite{BertschPratt,RiGy}.

Data~\cite{data} from BNL-RHIC, and also from the CERN-SPS, represents
a big ``puzzle''~\cite{Gyulassy:2001zv} in that the duration of particle
emission, as well as the longitudinal homogeneity length, appear rather short.
In fact, within the framework of relativistic perfect fluid dynamics,
they seem incompatible with even a mere cross-over~\cite{RiGy,ziesche}.
Moreover, studies
of hadronic rescattering following hadronization of a QGP fluid
indicated that one should always observe a long emission duration for hadrons
if they are formed at the confinement temperature $T_c\simeq
180\pm20$~MeV in equilibrium~\cite{Soff:2000eh}. 
This is because the lifetime of the particle source
is enhanced by dissipative effects. 

In these notes, I compute the corrections to the single-inclusive pion
distribution and to the two-pion correlation function due to
deviations from perfect equilibrium. I use a simple
schematic model which allows for an analytical treatment:
one-dimensional (longitudinal) boostinvariant expansion; of course, I
must focus on the longitudinal ``radius'' $R_\parallel$ then. The
analytical expression for the longitudinal homogeneity length~(\ref{Rl})
with first-order non-equilibrium corrections is a new result (to my knowledge),
generalizing the expression of Makhlin-Sinyukov~\cite{Makhlin:1987gm}
which applies to perfect
local equilibrium. Based on those results, I discuss why HBT correlation
functions are much more sensitive to the late, dissipative stages of the
evolution than the single-inclusive distributions (in agreement
with quantitative kinetic theory solutions~\cite{Soff:2000eh}): the
first-order theory applied here predicts that the correlation function
at small relative momentum of the pions can not in fact ``freeze out''.
This supports the expectation that one should observe a long duration
of emission if the post-QGP evolution starts with a hot ($T\simeq T_c$)
and dense hadron fluid close to equilibrium.
Hence, the way to go might be: start low.
A phase transition away from equilibrium can produce cool hadrons with
$T<T_c$, and induce large gradients of the temperature and velocity fields.
See the last section.

I do not review here the extensive literature on HBT interferometry for nuclear
collisions. Recent studies devoted specifically to effects from a confinement
transition in $Au+Au$ at RHIC include computations using 
\begin{enumerate}
\item ideal hydrodynamics~\cite{RiGy,ziesche,Schlei,KH_HBT,hirano},
\item ideal hydrodynamics up to hadronization followed by microscopic
hadron kinetic theory~\cite{Soff:2000eh},
\item parton kinetic theory with~\cite{lin} or without~\cite{molnar}
hadronic rescattering,
\item a QGP evaporation model without interactions among
hadrons~\cite{larry_sandra}.
\end{enumerate}
Those papers can also be used to trace earlier literature.

\section{First-order deviation from equilibrium}
A collection of many particles in a large, static volume approaches a steady
distribution in momentum space at times much larger than the
microscopic relaxation time for collisions $\tau_c$. This is the equilibrium
distribution $f_{\rm eq}(E)$, which is either a Bose or a Fermi distribution,
depending on the nature of the particles. The temperature $T$
characterizes how rapidly the probability to find a particle with energy
between $E$ and $E+dE$ falls off as $E$ increases. In equilibrium, $T$ is
constant throughout the entire volume.

More generally, one can think of situations where $T$ varies in space.
That is, on scales much larger than the interparticle distance but much
smaller than the total volume, the distribution of particle energies is
still described approximately by the equilibrium distribution. If this is not
the same distribution everywhere in space, one expects hydrodynamic flow to
develop between those regions.

In the absence of a uniform temperature and flow velocity field (global
equilibrium), there must also be local corrections to the equilibrium
distribution. They can be obtained from the Boltzmann
equation~\cite{Danielewicz:ww}
\be \label{BoltzEq}
p\cdot \partial f(x^\mu,p^\mu) = {\cal C}[f]~.
\ee
The collision kernel on the right-hand-side is a functional of the
distribution function $f(x^\mu,p^\mu)$. It vanishes for $f(x^\mu,p^\mu)=
f_{\rm eq}(E;T(x^\mu))$. However, the left-hand-side does not vanish if
$f(x^\mu,p^\mu)=f_{\rm eq}(E;T(x^\mu))$ with a non-uniform temperature field
$T(x)$ because of the presence of gradients. Therefore, $f_{\rm eq}
(E;T(x^\mu))$ is not a solution of the Boltzmann equation~(\ref{BoltzEq}).
Expanding the collision kernel
to first order in deviations from equilibrium, one can define the relaxation
time for collisions via
\be\label{tauc}
\frac{E}{\tau_c} = \frac{\delta {\cal C}}{\delta f}\Big|_{f_{\rm eq}}~.
\ee
Note that $\tau_c$ depends on both momentum and space-time, since
$\delta {\cal C}/\delta f$ is evaluated at $f_{\rm eq}(x^\mu,p^\mu)$.
For simplicity, in what follows I neglect the dependence on energy-momentum 
(more precisely, that on rapidity; $p_t$ dependence is allowed).

The local distribution function is then given by
\be
f(x^\mu,p^\mu) = 
   f_{\rm eq}(E(x^\mu);T(x^\mu)) + \delta f(E(x^\mu);T(x^\mu))
\equiv
   f_{\rm eq}(E(x^\mu);T(x^\mu)) \left[1+ 
      \widetilde{\delta f}(E(x^\mu);T(x^\mu))\right]~,
\ee
where $E(x^\mu)\equiv p\cdot u(x^\mu)$ is the energy of the particles measured
in the local rest frame. The Boltzmann equation then reads
\be
p\cdot\partial\left( f_{\rm eq}+\delta f\right) = 
    {\cal C}\left[f_{\rm eq}+\delta f\right] \simeq 
    {\cal C}\left[f_{\rm eq}\right]+
       \delta f \frac{\delta {\cal C}}{\delta f}\Big|_{f_{\rm eq}}~.
\ee
The first term on the right-hand-side vanishes by definition of $f_{\rm eq}$,
while the second term on the left-hand-side can be dropped if considering only
corrections to first order in gradients~\cite{Danielewicz:ww}.
Thus, using~(\ref{tauc}),
\be
\delta f = \frac{\tau_c}{p\cdot u} p\cdot\partial f_{\rm eq}~.
\ee
In what follows, I focus on the limit $m_t^2=p_t^2+m^2 \gg T^2$,
i.e.\ heavy and/or high-momentum particles, such that the quantum mechanical
distribution functions can be approximated by the classical Boltzmann
distribution. I also assume that no conserved
charges such as baryon number are present; in this case viscous corrections to
the energy-momentum tensor obtained from $\delta f$ vanish in the frame where
$u^\mu=(1,\vec{0})$, which therefore corresponds to the Landau-Lifshitz
definition of the local rest frame. Then,
\be \label{df_tilde}
\widetilde{\delta f}(x^\mu) = 
\frac{\tau_c}{p\cdot u(x^\mu)} \, p\cdot\partial\,
\frac{p\cdot u(x^\mu)}{T(x^\mu)}~. \label{FO_df}
\ee
Below, I will apply~(\ref{FO_df}) to compute corrections to the
two-particle correlation function. The Israel-Stewart second order
theory of imperfect fluids in principle
represents a more satisfactory, yet much more involved approach~\cite{azw}.

\section{Single-particle distribution}
The single-inclusive distribution of particles in a fluid,
measured on some space-time hypersurface $\sigma^\mu$
is given by~\cite{cooperfrye}
\be
\widetilde{N}(p) \equiv \frac{dN}{d^2p_t dy} =
\int d\sigma\cdot p \, f(p\cdot u)~.
\ee
Consider a fluid with infinite extent in the transverse directions,
and with space-like hypersurfaces of homogeneity.
In particular, for longitudinal scaling flow\footnote{This simply means
that the
flow rapidity $\eta_f={\rm Artanh}\,v_x$ equals the space-time
rapidity $\eta={\rm Artanh}\,(x/t)$ everywhere in the forward light-cone,
where $x$ denotes the longitudinal direction.} and vanishing
transverse flow velocity, those are surfaces of fixed proper time
$\tau=\sqrt{t^2-x^2}$, which is invariant under Lorentz boosts. Then,
\be
d\sigma\cdot p = d^2r\, d\eta\, \tau m_t \cosh(y-\eta)~.
\label{dsigp}
\ee
Thus,
\bea
\widetilde{N}(p) &=&
 \int d^2r\, d\eta\, \tau m_t \cosh(y-\eta) \left\{
 f_{\rm eq}\left(p\cdot u\right) + \delta f\left(p\cdot u\right)\right\}
  \nonumber\\
&=& \int d^2r\, d\eta\, \tau m_t \cosh(y-\eta) f_{\rm eq}\left(p\cdot u\right)
 \left\{ 1+ \widetilde{\delta f}\left(p\cdot u\right)\right\}~.
 \label{sing_incl}
\eea
At large $m_t/T$ the Bose or Fermi distributions approach a Boltzmann
distribution. Here $T$ denotes the temperature on the hypersurface
specified by $\tau$; for
simplicity, let it be independent of the rapidity of the flow.
Moreover, in this limit the single-inclusive distribution can
be evaluated by a saddle-point integration,
\be
f_{\rm eq}\left(p\cdot u\right) \simeq e^{-p\cdot u/T} =
e^{-m_t\cosh(y-\eta)/T} \rightarrow \sqrt{\frac{2\pi T}{m_t}}\,
\delta(y-\eta) \, e^{-m_t/T}~.
\label{saddle}
\ee
In other words, evaluate the integral over the flow rapidity $\eta$ by
expanding (in the exponential)
$p\cdot u=m_t\cosh\theta\simeq m_t(1+\theta^2/2)$ up to
first nontrivial order in $\theta\equiv y-\eta$,
and set $y=\eta$ throughout the rest of
the integrand in~(\ref{sing_incl}); this reduces the integral over $\eta$
to a Gaussian integral.

The saddle-point approximation also greatly simplifies $\widetilde{\delta f}$:
\bea
p\cdot\partial \frac{p\cdot u(\sigma)}{T(\sigma)} &=& 
\frac{1}{T(\sigma)} p\cdot\partial p\cdot u(\sigma) +
p\cdot u(\sigma) p\cdot\partial \frac{1}{T} =
- \frac{m_t^2}{\tau T}\sinh^2\theta +m_t^2\cosh\theta \left[
  \cosh\theta\partial_\tau+\frac{1}{\tau}\sinh\theta\partial_\eta\right]
  \frac{1}{T(\sigma)} \nonumber\\
& & \quad\rightarrow m_t^2\partial_\tau \frac{1}{T}~,
\eea
and so
\be
\widetilde{\delta f} (p\cdot u) \to \tau_c m_t \partial_\tau \frac{1}{T}~.
\ee
The integral over $d^2r$ is trivial, since we assumed that neither the
temperature $T(\sigma)$ nor the flow velocity $u^\mu(\sigma)$
depend on $r$; thus, $\int
d^2r$ equals the transverse area $S_t=\pi R^2$. We obtain
\be
\widetilde{N}(p) = S_t \, \tau \sqrt{2\pi m_t T}\, e^{-m_t/T}\left\{
1-\frac{\tau_c}{\tau} \frac{m_t}{T}\frac{\partial \log T}{\partial\log\tau}
\right\}~.
\ee
To be more explicit, we may for example assume that the temperature prior to
freeze-out decreases like a power-law in $1/\tau$:
\be \label{T_tau_law}
T \sim \frac{1}{\tau^\gamma}~.
\ee
For ideal isentropic expansion the entropy in a comoving volume element is
conserved, which for ultrarelativistic particles ($m\ll T$)
leads to $\gamma=1/3$. On the other hand, isoergic expansion
is perhaps more realistic near freeze-out; in that case the energy per
comoving volume element is conserved, which gives $\gamma=1/4$. Then,
\be
\widetilde{N}(p) = S_t \, \tau \sqrt{2\pi T m_t} \, e^{-m_t/T}\left\{
1+\gamma\frac{\tau_c}{\tau} \frac{m_t}{T}\right\}~.
\label{single}
\ee
The leading term in eq.~(\ref{single}), corresponding to $\tau_c\to0$, is the
result for an equilibrium distribution of the particles.
The second term in the curly brackets is due to
corrections from local equilibrium, proportional to the microscopic relaxation
time $\tau_c$ relative to the expansion time $\tau$. 
It gives us a hint about the meaning of ``freeze-out''. The hadrons measured
in the detector are cold $T=0$ hadrons; they do not stop interacting suddenly
at some temperature\footnote{For example, for a chiral symmetry which
is broken spontaneously (by choice of a vacuum) but not explicitly,
at low $T$ corrections to the pressure of non-interacting pions are determined
by chiral perturbation theory: $\sim(T/f_\pi)^4\,p_{\rm id}$.}. However,
as $T$ decreases in the course
of the evolution, corrections to $f_{\rm eq}$ become more and more important.
The more of them I include in
my computation, the lower I can go in $T$. With the full collision kernel
I could go down smoothly to $T=0$~\cite{Bass:1999tu} without
introducing an abrupt ``decoupling''. The important point to realize though
is that as corrections to equilibrium grow, they tend to ``stabilize'' the
single-inclusive distribution: from eq.~(\ref{single}) one sees that the ideal
$\widetilde{N}(p)$ at any fixed $m_t$ decreases with
$T$, but the correction term acts to slow down that evolution.
Eventually, at some time $\tau$ or temperature $T$,
the correction terms ``freeze'' $\widetilde{N}(p)$,
i.e.\ it stops evolving any further despite the ongoing approach of $T$
to zero. I can view that as the ``freeze-out'' temperature. 

What this discussion points at, as well, is that ``freeze-out'' is not
universal but depends on the {\em observable} under consideration. This is
because corrections from deviation from perfect equilibrium are not the
same for all observables. We shall return to that below.

The correction to first order in the relaxation time to equilibrium, $\tau_c$,
does not make for a quantitative description of the freeze-out process.
Nevertheless, to get a rough idea, one might ask when $d\widetilde{N}(p)/d\tau
=0$. This leads to
\be
\dot\tau_c = 
1+\frac{\gamma m_t}{T} \frac{\tau_c}{\tau}
  -\frac{T}{m_t} \frac{1-\gamma/2}{\gamma}
  -\frac{\gamma}{2}\frac{\tau_c}{\tau}
=1+\frac{\gamma m_t}{T_0}\left(\frac{\tau}{\tau_0}\right)^\gamma
  \frac{\tau_c}{\tau}
  -\frac{T_0}{m_t} \frac{1-\gamma/2}{\gamma}\left(\frac{\tau}{\tau_0}\right)
  ^{-\gamma}-\frac{\gamma}{2}\frac{\tau_c}{\tau}~. \label{tauc_tau}
\ee
Here, $\tau_0$ is the time when corrections to perfect equilibrium set in.
This could be as early as hadronization, or any time after that.
$T_0$ denotes the temperature at that time ($m_t/T_0$
must be large enough so that $\dot\tau_c>0$).

At $\tau_0$ we may, for example, be dealing with a perfect fluid, i.e.\
$\tau_c(\tau_0)=0$. As long as $\tau_c/\tau$ remains negligibly small,
and $\tau/\tau_0$ is not too big,
the solution of~(\ref{tauc_tau}) is approximately given by
\be
\tau_c = \tau-\tau_0-\frac{T_0\tau_0}{m_t} \frac{1-\gamma/2}{\gamma(1-\gamma)}
\left[\left(\frac{\tau}{\tau_0}\right)^{1-\gamma}-1\right]~.  \label{tauc_lin}
\ee
Thus, $\tau_c$ should grow approximately linearly with the expansion time.
As $\tau_c$ becomes larger and comparable to $\tau$,
its growth must speed up in order that $\widetilde{N}(p)$ remains
``frozen'' (also, if it is
non-negligible already at hadronization, i.e.\ if the hadron fluid is rather
viscous from the start). To see this, keep only terms proportional
to $\tau_c$ on the right-hand side of~(\ref{tauc_tau}), which leads to
\be
\frac{\tau_c(\tau)}{\tau_c(\tau_0)} = \left(\frac{\tau}{\tau_0}\right)
^{-\gamma/2}\, \exp\left(\frac{m_t}{T_0}\left[\left(\frac{\tau}
{\tau_0}\right)^\gamma-1\right]\right)
= \sqrt{\frac{T}{T_0}}\, e^{m_t/T-m_t/T_0}~.  \label{tauc_exp}
\ee
Clearly, when the rapid growth kicks in, this signals the break-down of
the first-order theory and that corrections of
higher order in $\tau_c$ also become important.

So, to summarize, the basic picture is as follows. A hadron ``fluid''
is produced by the decay of the deconfined state. In theory, it may
evolve as an ideal fluid for some time, if the relaxation time to
equilibrium is extremely short. Eventually, as the fluid becomes
more dilute, the relaxation time must grow: at first rather slowly, then
turning to a rapid non-linear growth. Alternatively, the relaxation
time may be non-negligible already at hadronization, and continue to grow
rapidly as the fluid expands. In either case, as soon as the growth rate of
$\tau_c$ satisfies eq.~(\ref{tauc_tau}), the single-particle distribution
doesn't change any more, it ``freezes''. Smoothly, the fluid cools to
$T=0$\footnote{At some point number-changing reactions must also freeze.
One then has to introduce a continuity equation for the number-current,
which will lead to a chemical potential~\cite{teaney_chem}.}.

\section{Two-particle correlation function}
The two-particle correlation function for identical particles is given by
\bea
R(p_1,p_2)&\equiv& \widetilde{N}(p_1) \widetilde{N}(p_2) \left[ C_2(p_1,p_2)-
        1\right]\nonumber\\
 &=& \mbox{Re} \int d\sigma_1\cdot K d\sigma_2\cdot K \, \exp\,
i(\sigma_1-\sigma_2)\cdot(p_1-p_2) \nonumber\\
& & \quad \times 
\left\{ f_{\rm eq}\left(u_1\cdot K\right)
      + \delta f\left(u_1\cdot K\right) \right\}
\left\{ f_{\rm eq}\left(u_2\cdot K\right)
      + \delta f\left(u_2\cdot K\right) \right\}~,
\label{R_p1_p2}
\eea
with $K^\mu=(p_1^\mu+p_2^\mu)/2$.
I consider particle pairs emitted at the same azimuthal angle and
with the same transverse momentum, $\vec{p}_{t,1}=\vec{p}_{t,2}$. This is no
severe restriction as for 1+1d longitudinal expansion
only differences in the longitudinal momenta of the emitted
particles matter. Introducing the variables $\alpha=y_2-y_1$ for the
relative rapidity of the emitted particles and $\theta_1=y_1-\eta_1$,
$\theta_2=y_2-\eta_2$ for their rapidities relative to the emitting fluid
elements, the phase factor becomes
\bea
e^{i(\sigma_1-\sigma_2)\cdot(p_1-p_2)} &=&
e^{i m_t\tau(\cosh\theta_1-\cosh(y_2-\eta_1)+\cosh\theta_2-\cosh(y_1-\eta_2))}=
e^{i m_t\tau((1-\cosh\alpha)(\cosh\theta_1+\cosh\theta_2)
             -\sinh\alpha(\sinh\theta_1-\sinh\theta_2))}\nonumber\\
&=& 
e^{i \alpha m_t\tau( \theta_2-\theta_1 - \alpha - 
                     \alpha(\theta_1^2+\theta_2^2)/4) +
       {\cal O}(\alpha^3,\theta_{1,2}^3)}~.
\eea
In anticipation of the saddle-point integration over the flow rapidities to be
performed below, I expanded the exponent up to second order in $\theta_1$ and
$\theta_2$. Also, I shall be interested mainly in the behavior of the
correlation function at small relative rapidity and so expanded up to second
order in $\alpha$.

The same manipulations lead to
\bea
d\sigma_1\cdot K &=& \frac{\tau m_t}{2}\left[\left(1+\cosh\alpha\right)
\cosh\theta_1+\sinh\alpha\sinh\theta_1\right]d\theta_1\,d^2r_1 \nonumber\\
&=&  \frac{\tau m_t}{2}\left[\left(1+\frac{\alpha^2}{4}+\frac{\alpha}{2}\right)
 e^{\theta_1} +\left(1+\frac{\alpha^2}{4}-\frac{\alpha}{2}\right) e^{-\theta_1}
\right]d\theta_1\,d^2r_1~,\\
f_{\rm eq}\left(u_1\cdot K\right) &=&
e^{- \frac{m_t}{2T}\left(\cosh\theta_1+\cosh\left(y_2-\eta_1\right)\right)}
= e^{-\frac{m_t}{2T}\left[\left(1+\cosh\alpha\right)\cosh\theta_1
                           +\sinh\alpha\sinh\theta_1\right]}\nonumber\\
&=& 
e^{-\frac{m_t}{2T}\left[\left(2+\alpha^2/2\right)\left(1+\theta_1^2/2\right)
  +\alpha\theta_1\right] + {\cal O}(\alpha^3,\theta_{1}^3)}~.
\eea
The expressions for $d\sigma_2\cdot K$ and $f_{\rm eq}(u_2\cdot K)$ can be
obtained by substituting $\theta_1\to\theta_2$ and $\alpha\to-\alpha$.
Integrating over the rapidity $\theta_1$ gives
\bea
& & e^{-i\alpha^2 m_t\tau/2}
\int\limits_{-\infty}^\infty d\theta_1 \,
  e^{-i\alpha m_t\tau(\theta_1+\alpha\theta_1^2/4)}
  e^{-\frac{m_t}{2T} (\alpha\theta_1 +(1+\alpha^2/4)\theta_1^2)}
\left[\left(1+\frac{\alpha^2}{4}+\frac{\alpha}{2}\right)
 e^{\theta_1} +\left(1+\frac{\alpha^2}{4}-\frac{\alpha}{2}\right) e^{-\theta_1}
\right]\nonumber\\
&\simeq& 
 \sqrt{\frac{2\pi T}{m_t}\left(1-\frac{\alpha^2}{4}\left(1+2i\tau T\right)\right)}
\,\, e^{-\frac{m_t}{T}\frac{\alpha^2}{2}\tau^2 T^2} 
\left(2+\frac{\alpha^2}{4}\right)\cos(\alpha\tau T)~. \label{eq35}
\eea
On the right-hand side, I performed the following approximations.
I expanded the arguments of the exponentials assuming
$\alpha\ll1$ but $\alpha m_t/T$ and $\alpha\tau T$ of order one. 
I kept only terms up to ${\cal O}(1)$ in this approximation
scheme. Nevertheless, I dropped a term $(i/4)\alpha^4(m_t/T)(\tau T)^3=
{\cal O}(1)$ since I will mainly be interested in the curvature of
the correlation function at $\alpha=0$.
Since~(\ref{eq35}) is symmetric in $\alpha\to-\alpha$, the integral over
$\theta_2$ gives the same result. 

Next, I need $\widetilde{\delta f}(u_1\cdot K)$ in
the limit $\theta_1\to0$, up to second order in $\alpha$. The result is
\be
\widetilde{\delta f}(u_1\cdot K) =
\tau_c m_t\left\{\left[\left(1+\frac{\alpha^2}{4}\right)\partial_\tau
+\frac{\alpha}{2\tau}\partial_{y_1}\right]\frac{1}{T}-\frac{\alpha^2}{4T\tau}
\right\}~.
\ee
The result for $\widetilde{\delta f}(u_2\cdot K)$ is the same,
except that the derivative of $1/T$ with respect to rapidity is evaluated at
$y_2$, respectively, and $\alpha\to-\alpha$. Then,
\bea
R(m_t,\alpha) &=& \int d^2r_1 d^2r_2 \, \tau^2 2\pi T m_t \, e^{-2m_t/T}
\cos^2(\alpha\tau T)\, e^{-m_t\alpha^2\tau^2 T}\nonumber\\
& &\hspace{-1.5cm} \times 
\left\{1+\tau_c m_t
\left\{\left[\left(1+\frac{\alpha^2}{4}\right)\partial_\tau
+\frac{\alpha}{2\tau}\partial_{y_1}\right]\frac{1}{T}-\frac{\alpha^2}{4T\tau}
\right\}\right\}
\left\{1+\tau_c m_t
\left\{\left[\left(1+\frac{\alpha^2}{4}\right)\partial_\tau
-\frac{\alpha}{2\tau}\partial_{y_2}\right]\frac{1}{T}-\frac{\alpha^2}{4T\tau}
\right\}\right\}~.
\eea
For any function $f(y)$, to leading order in $\alpha=y_2-y_1$
\be
f'(y_2) = f'(y_1)+\alpha f''(y_1)~.
\ee
For $f(y)=1/T(y)$ then,
\be
\partial_{y_2}\frac{1}{T} = -\frac{1}{T^2}\partial_{y_1}T +
\alpha\left( -\frac{1}{T^2}\partial^2_{y_1}T + \frac{2}{T^3}\left(
   \partial_{y_1}T\right)^2\right) =
\frac{2\alpha}{T}\left(\partial_{y_1}\log T\right)^2 
-\frac{1}{T}\partial_{y_1}\log T-
\frac{\alpha}{T^2}\partial^2_{y_1}T~.
\ee
On the right-hand-side, $T$ is evaluated at $y_1$.
Integrating over the transverse area, and dividing by the product of the
single-particle distributions leads to
\bea
C_2(m_t,\alpha)-1 &=& 
\cos^2(\alpha\tau T)\, e^{-m_t \alpha^2\tau^2 T}\nonumber\\
&\times&
\left\{ 1-\frac{\tau_c}{\tau}\frac{m_t}{T}\frac{\alpha^2}{2}\left[1
+\left<\frac{\partial\log T}{\partial\log\tau}\right>
-\frac{1}{T}\left<\frac{\partial^2 T}{\partial y_1^2}\right>
+2\left<\left(\frac{\partial\log T}{\partial y_1}\right)^2\right>
\right]\right\}~. \label{cor_func}
\eea
Here, $\langle\cdot\rangle$ refer to averages over the transverse plane as
well as over events. For consistency, only terms to first order in $\tau_c$
have been kept.

We can define a longitudinal homogeneity
length as the curvature of the correlation function at $\alpha=0$:
\be
R^2_\parallel = -\frac{1}{m_t^2} \frac{1}{2} \frac{\partial^2 C_2(\alpha)}
{\partial\alpha^2}\Big|_{\alpha=0} = \frac{\tau^2 T^2}{m_t^2}+
\frac{\tau^2 T}{m_t}
+ \frac{\tau_c}{\tau} \frac{1}{2Tm_t} \left( 1-\gamma
-\frac{1}{T}\left<\frac{\partial^2 T}{\partial y_1^2}\right>
+2\left<\left(\frac{\partial\log T}{\partial y_1}\right)^2\right>\right)~.
\label{Rl2}
\ee
The $1/m_t^2$ prefactor arises because one should in fact take the derivative
with respect to the longitudinal momentum difference; for small relative
rapidity $\alpha$, and in the logitudinally comoving frame,
$p_\parallel/m_t = \tanh\alpha\simeq\alpha$. The first term in~(\ref{Rl2})
is subleading in $T/m_t\ll1$ and will be dropped. Then,
\be
R_\parallel = \tau \sqrt{\frac{T}{m_t}} \sqrt{1+
\frac{\tau_c}{\tau} \frac{1-\gamma-\zeta_T +2\Delta_T^2}{2\tau^2 T^2}}~,
\label{Rl}
\ee
with
\be
\Delta_T^2 \equiv \left<\left(\frac{\partial\log T}{\partial y_1}\right)^2
\right>~,~
\zeta_T = \frac{1}{T}\left<\frac{\partial^2 T}{\partial y_1^2}\right>~.
\ee
Deviations from equilibrium
manifest themselves in the second term under the square root.
They do not distort the scaling of $R_\parallel$ with $1/\sqrt{m_t}$
predicted from a local equilibrium distribution~\cite{Makhlin:1987gm}
(regarding scaling of the $out$ and $side$ radii see~\cite{BLfit}).
Note, however, the ``wrong sign'' of the correction relative to the
leading term, to which we shall return below in eq.~(\ref{tauc_Rl}).

$\Delta_T^2$ is the mean-square fluctuation of the temperature\footnote{The
{\em average} temperature field was of course assumed to be rapidity
independent, $\langle\partial T/\partial y\rangle=0$.}, divided by
the typical rapidity scale $\Delta y$ on which it occurs; it measures
fluctuations about the average homogeneous particle source\footnote{These
are analogous to the
temperature fluctuations of the cosmic microwave background measured by
COBE~\cite{COBE}, of order $\Delta T/T\simeq10^{-5}$.}. 
Only fluctuations on scales $\Delta y$
smaller than the relative rapidity $\alpha$ at
which the correlation function $C_2(\alpha)$ is being 
probed\footnote{The single-inclusive distribution~(\ref{single})
is not sensitive to $\Delta_T$. That is, of course, due to the fact that it
measures the temperature at only one rapidity but not correlations
of $T$ between two rapidities $y_2$ and $y_1$. One might be tempted
to measure the single-particle distribution in very narrow bins in
rapidity, and in individual events,
such as to trace any local disturbance of a boost invariant temperature
distribution. This is not possible, however, because the
saddle-point~(\ref{saddle}) of the distribution function has a width
$\sqrt{2\pi T/m_t}$ in rapidity, which can not be less than $\sim1$ for
values of $m_t$ where the application of the collective hydrodynamic
theory makes sense. Thus, temperature fluctuations on scales $\Delta y\ll1$
get washed out.} matter. Fluctuations on
larger scales do not matter as in that case both particles are emitted from
the same rapidity ``element'' and at the same $T$, so $\Delta T=0$. In any
case, such rapidity fluctuations of $T$ 
can only give a significant correction to $R_\parallel$ if $\Delta y\ll1$ and
if $\Delta T/T$ is of order unity. In other words, after the transition, there
should be regions of longitudinal extension $\sim1$~fm (which at time
$\tau\simeq10$~fm corresponds to a rapidity interval of 0.1) where $T$ is
about equal the
ordinary hydrodynamic freeze-out temperature for a homogeneous fluid,
separated by cold regions of similar size with $T\sim0$ (see
also~\cite{Heiselberg:1998cv}).
Nevertheless, temperature fluctutations of order $\Delta T/T\sim1$ appear
unrealistically large.
Moreover, experimentally one usually measures correlations of identical
pions, which are the most abundant hadron species. A significant part of the
pion yield is commonly believed to originate from decays of resonances ($\rho$,
$\eta$, $\Delta$, ...) after freeze-out. Those resonance decays
smear out any temperature
fluctuations on scales $\Delta y\ll1$, even if those were
present before the resonance decays. Therefore,
realistically it seems that the correction proportional to
$\Delta_T^2$ in eq.~(\ref{Rl}), which would tend to enhance $R_\parallel$,
is rather small.

There is also an increase ($1-\gamma>0$) of $R_\parallel$ by an amount
depending on the temperature gradient in time direction.
If that gradient $\gamma$ is small, which corresponds to a rather viscous
fluid, the increase is largest. If the temperature drops steeply, $\gamma$
is larger and so the correction is smaller. In any case, by considering
either isentropic or isoergic expansion, i.e.\ conservation of either
the entropy or the energy in a comoving volume element, one obtains
$\gamma\simeq 1/4-1/3$, and so $1-\gamma={\cal O}(1)$.

The only term leading to a {\em reduction} of $R_\parallel$ is that
proportional to $\zeta_T$, the curvature of the temperature field in rapidity.
This is also obvious intuitively. Note that this curvature term
arises from the first-order correction to local equilibrium (it is proportional
to $\tau_c$). A nonvanishing curvature of $T$ does not contradict
the initial assumption that $\langle\partial T/\partial y\rangle=0$
at midrapidity. Moreover, one could still argue that $T$ is rapidity
independent only on scales probed by the single-inclusive distribution,
say $\Delta y\simeq1$, but does exhibit some local curvature on scales
$\Delta y\ll1$. Even so, it doesn't seem likely that the {\em average}
local curvature, averaged over the transverse plane and events, is big.

Therefore,
in summary, all of the corrections from eq.~(\ref{Rl}) proportional to $\tau_c$
should be rather small for the case of high-energy nuclear collisions, in
particular since they are down by a factor $1/(\tau T)^2$ relative to
perfect local equilibrium: for nuclear collisions, typically $\tau$ is on the
order of 10~fm and $T$ on the order of 100~MeV, so $\tau T\sim 5$. 

This brings us to the main point. From kinetic solutions for the evolution
of hadrons produced from a QGP fluid it was observed that the two-particle
correlation functions are sensitive to a ``cloud'' of late soft hadronic
interactions, which only minorly affect the single-inclusive
distribution~\cite{Bass:1999tu,Soff:2000eh}. We can understand qualitatively
why this is so. Above, I argued that deviations from local equilibrium grow
as the temperature $T$ decreases and eventually ``freeze'' the single-inclusive
distribution~(\ref{single}), despite the smoothly ongoing cooling
down to $T=0$.

The correlation function~(\ref{cor_func}) also exhibits a correction term
proportional to $\tau_c/\tau$. Note, however, that the correction vanishes
at small relative rapidity (or longitudinal momentum) of the pions,
$\alpha\to0$~! The characteristic relative rapidity over which $C_2$ falls off
is set by the exponential in eq.~(\ref{cor_func}), $\alpha^2\simeq 1/(m_t
\tau^2 T)$. Thus, the first-order correction in the curly brackets
in~(\ref{cor_func}) is of order $\tau_c/(\tau^3 T^2)$,
and can only become large when $\tau_c/\tau \simeq \tau^2 T^2$.
This is a large number for the case of nuclear collisions, and it
grows with time. This means that if we are dealing with a nearly perfect
hadron fluid ($\tau_c/\tau\ll1$)
with only linear growth of $\tau_c$ in time, as in
eq.~(\ref{tauc_lin}), then non-ideal corrections to $R_\parallel$
actually {\em decrease} with time. This is despite the fact that
they are sufficient to actually ``freeze'' the single-inclusive
distribution~! 

For $R_\parallel$ to ``freeze'', i.e.\ $dR_\parallel/d\tau=0$, the
relaxation time must evolve as
\be
\dot\tau_c = (1-\gamma)\frac{\tau_c}{\tau}-2\frac{2-\gamma}{1-\gamma}
T^2\tau^2~, \label{tauc_Rl}
\ee
where I dropped $\zeta_T$ and $\Delta_T$ for simplicity.
Thus, close to equilibrium, i.e.\ for small $\tau_c/\tau$, 
$R_\parallel$ can not ``freeze'' since the relaxation time would have to
decrease with time, which seems unphysical\footnote{At least, if
one has in mind typical kinetic estimates of relaxation times, see for
example~\cite{PPVW}. Near a critical point estimates of $\tau_c$
might be altered.}. 
Eq.~(\ref{tauc_Rl}) shows that there is no physical solution ($\dot
\tau_c\ge0$ for $\tau>\tau_0$) which leads
to ``freezing'' of $R_\parallel$ in a fluid near equilibrium, contrary
to $\widetilde{N}(p)$. That is because for $\dot\tau_c$ to stay
positive, the first term on the right-hand-side of eq.~(\ref{tauc_Rl})
must overwhelm the second one. However, the first term by itself asks for
a rather slow, nearly linear increase $\tau_c\sim\tau^{1-\gamma}$. In turn,
the negative term in~(\ref{tauc_Rl}) grows almost quadratically with
$\tau$, and so will eventually take over. The only way out is to
stabilize $R_\parallel$ through corrections of higher order in
$\tau_c$, which are not taken into account in~(\ref{Rl2}).
This motivates why the HBT radii are much more sensitive to the
dissipative evolution of the hadron ``fluid'' than the single-inclusive
distribution or quantities derived from it (flow~\cite{flow}).

The ``HBT puzzle'' can therefore be formulated as follows: if hadronization
leads to a hot and dense hadron fluid, $\tau_c/\tau$ is small
initially\footnote{Here, ``initial'' refers to the initial condition for
the hadrons, which is set by the hadronization.}. It will take a long
time until the two-particle correlation function ``freezes out'' because
$\tau_c$ first has to evolve slowly into the regime where
corrections of higher order in $\tau_c$
become important. Consequently, $R_\parallel$ will be large.
In fact, assuming that the initial temperature of the hadrons is equal to
the confinement temperature $T_c\simeq 180\pm20$~MeV already leads to
significant deviations of the computed HBT radii~\cite{Soff:2000eh,soff2}
to the data~\cite{data}. The theoretical calculations do not fail
very badly, but in a clear systematic fashion. While the computed ratio of the
{\sl out} and {\sl side} radii, $R_o/R_s$, increases with the
pair transverse mass $m_t$, the data shows it rather flat.
Also, the experimental $R_\parallel(m_t)$ does seem to roughly follow the
$1/\sqrt{m_t}$ behavior but is systematically
lower than the theory at all $m_t$, even for the
most ``optimistic'' assumption $T_c=160$~MeV.

\section{How it might work...}
If a slow growth of the relaxation time prevents $R_\parallel$ from
freezing quickly, how could one then obtain a small homogeneity
length~?

It clearly helps to start the final-state hadron kinetic evolution
at low temperature, and with big deviations from equilibrium.
The closer the inital condition for the hadronic evolution
is to the ``freeze-out'' condition, the better. The results
of~\cite{Soff:2000eh} already showed that the picture improves if
the initial temperature of the hadron fluid is lowered, but the use
of perfect-fluid dynamics as a hadronization scheme implies that
on the hadronization surface $\tau_c/\tau$ was small in those
studies. As discussed above, the QGP shouldn't hadronize
into an ideal fluid of hadrons, but into a very viscous one.

Employing ideal hydrodynamics to model the phase transition, the fluid of
hadrons starts out essentially at $T_c$. With $T_c\ge160$~MeV, one can not
describe the HBT data~\cite{data} very well~\cite{Soff:2000eh,soff2}:
the hadronic fluid is too dense and $\tau_c$ too small.
One should probably go lower in temperature,
which implies a non-equilibrium phase
transition. One way to obtain it is from parton cascade
models~\cite{lin,molnar} where one can tune the parton$\to$hadron
transition such as to yield a cool hadron fluid with
initial temperature $T_0$ below $T_c$ and with large velocity and
temperature gradients. (Increasing the parton-parton
cross section relative to lowest-order pQCD estimates
presumably does the job by delaying the transition into hadrons.)
This might be the reason why ref.~\cite{lin} apparently is able to fit
the radii at RHIC (in fact, $R_\parallel$ even comes out too small).
In field-theory language, in turn, one would have to consider
spinodal decomposition~\cite{scavenius} if the transition is first-order.
If it is a cross over or weak first-order transition,
the decay of a condensate which saturates the free energy of the deconfined
state at $T_c$~\cite{PLMdec} could be a successful approach:
while the deconfinement temperature sets the scale for the effective potential,
the reheating temperature of the hadrons is determined by the decay
process of the condensate and by the overall expansion, and so
is quite likely less than $T_c$\footnote{This is in close analogy to reheating
after inflation. Integrating out hard fluctuations determines
the temperature scale in the effective potential for the zero mode
(condensate) of the inflaton field~\cite{linde}. The reheating
temperature, i.e.\ the temperature of the particles produced from the decay
of that condensate, is however determined dynamically by the decay process.}.

In summary, the pion correlation function requires much smaller
relaxation rate than the single-particle distribution in order to ``freeze''.
Therefore, HBT
interferometry of the QGP ``ashes''~\cite{BertschPratt} might reveal some
of the bulk properties of the QGP hadronization and its non-equilibrium
nature. The first-order theory applied here is not capable of quantitative
predictions but it clearly points at the fact that ``freeze-out'' of
two-particle correlations at small relative momentum is tightly related
to the {\em dynamics} of the confinement transition.
\acknowledgements
I have benefitted from discussions with Larry McLerran, Dirk Rischke
and Derek Teaney.

\end{document}